\documentclass[notitlepage,a4paper,aps,prd,onecolumn,superscriptaddress,nofootinbib,groupedaddress,longbibliography]{revtex4-2}
\usepackage{amsmath}
\usepackage{amsfonts}
\usepackage{amssymb}
\usepackage{accents}
\usepackage[utf8]{inputenc}
\usepackage[T1]{fontenc}
\usepackage[colorlinks=true]{hyperref}
\usepackage{tikz}
\usetikzlibrary{arrows.meta,external}
\newcommand{\order}[2]{\accentset{#2}{#1}}
\newcommand{\lc}[1]{\accentset{\circ}{#1}}
\newcommand{\dd}{\mathrm{d}}
\newcommand{\mL}{\mathcal{L}}

\begin{document}

\title{Parametrized post-Newtonian limit of generalized scalar-nonmetricity theories of gravity}

\author{Kai Flathmann}
\email{kai.flathmann@uni-oldenburg.de}
\affiliation{Institut f\"ur Physik, Universit\"at Oldenburg, 26111 Oldenburg, Germany}

\author{Manuel Hohmann}
\email{manuel.hohmann@ut.ee}
\affiliation{Laboratory of Theoretical Physics, Institute of Physics, University of Tartu, W. Ostwaldi 1, 50411 Tartu, Estonia}

\begin{abstract}
In this article we calculate the post-Newtonian limit of a general class of scalar-nonmetricity theories of gravity. The action is assumed to be a free function of the nonmetricity scalar, the kinetic term of the scalar field, two derivative couplings and the scalar field itself. We use the parametrized post-Newtonian formalism to solve the arising field equations for the case of a massless scalar field in order to compare several subclasses of this theory to solar system observations. In particular, we find several classes of theories which are indistinguishable from general relativity on the post-Newtonian level and therefore, should be studied further. Most remarkably, we find that this is the generic case, while a post-Newtonian limit that deviates from general relativity occurs only for a particular coupling between the scalar field and nonmetricity.
\end{abstract}

\maketitle

%\tableofcontents

%%%%%%%%%%%%%%%%%%%%%%%%%%%%%%%%%%%%%%%%%%%%%%%%%%%%%%%%%%%%%%%%%%%%%%%%%%%%%%%%%%%%%
\section{Introduction}\label{sec:intro}
Being contested in numerous experiments during the past century, general relativity is still the best gravitational theory describing observations in our universe. However, by fixing the mediator of gravity through the Ricci scalar as curvature two equivalent possibilities are overlooked~\cite{BeltranJimenez:2019tjy}. The first alternative ascribes gravity to the dynamics of the tetrad via the torsion scalar. This equivalent formulation of general relativity is called the Teleparallel Equivalent of General Relativity (TEGR)~\cite{Aldrovandi:2013wha,Bahamonde:2021gfp}. If we assume curvature and torsion as being zero and simultaneously nonmetricity as nonvanishing, we can construct the Symmetric Teleparallel Equivalent of General Relativity (STEGR)~\cite{Nester:1998mp,Adak:2005cd,Adak:2008gd,Mol:2014ooa,Adak:2018vzk,Jimenez:2019yyx,DAmbrosio:2020nqu,DAmbrosio:2020njs}. Another possibility invokes both torsion and nonmetricity~\cite{Jimenez:2019ghw,Boehmer:2021aji}. However, even though all of these formulations are equivalent, generalizations thereof differ from each other.

A vast number of modifications of general relativity and its alternative formulations in terms of torsion and nonmetricity mentioned above has been developed~\cite{CANTATA:2021ktz}. The main motivation for studying such theories comes from tensions between general relativity and current observations in cosmology, such as different measurements of the Hubble expansion rate~\cite{Planck:2018vyg}. These observations hint towards physics beyond the so-called $\Lambda$CDM model, which aims to describe the universe using general relativity, a cosmological constant \(\Lambda\) and cold dark matter (CDM). Among the most common modifications studied to address these tensions are generalizations of the action functional to a free function of the aforementioned scalar invariants of curvature, torsion or nonmetricity, giving rise to the so-called $f(R)$, $f(T)$ and $f(Q)$ classes of gravity theories~\cite{Sotiriou:2008rp,Bengochea:2008gz,Linder:2010py,BeltranJimenez:2017tkd,BeltranJimenez:2018vdo}. Another, related type of modifications is obtained by including an additional scalar field in the theory, which couples non-minimally to the geometric quantities which mediate the gravitational interaction, and can thus itself be regarded as a mediator of gravity. This type of modifications gives rise to scalar-curvature, scalar-torsion and scalar-nonmetricity theories of gravity~\cite{Fujii:2003pa,Geng:2011aj,Hohmann:2018rwf,Hohmann:2018vle,Hohmann:2018dqh,Hohmann:2018ijr,Jarv:2018bgs,Runkla:2018xrv}.

In order to be considered as a viable theory of gravity, any of the aforementioned modifications must not only address the observational tensions in cosmology, but also be in agreement with numerous precision observations of gravitational waves~\cite{Yagi:2016jml,Gair:2012nm,Carson:2020rea} and gravity on stellar or solar system scales~\cite{Will:2014kxa}. The latter can comprehensively be studied using the parametrized post-Newtonian (PPN) formalism~\cite{Will:1993ns,Will:2018bme}, which characterizes any metric theory of gravity by ten (usually constant) parameters. Their values predicted by any given theory of gravity can then be compared to their experimentally measured values, giving constraints on the considered theory or its parameters.

In this article we make use of the PPN formalism in order to derive the post-Newtonian limit of a general class of scalar-nonmetricity theories of gravity, which generalizes the originally proposed class~\cite{Jarv:2018bgs}, following a similar idea as applied in scalar-torsion gravity~\cite{Hohmann:2018dqh}, and allowing for a gravitational action defined by an arbitrary function of the nonmetricity scalar, two non-minimal coupling terms, the scalar field and its kinetic energy, and which we will therefore denote $\mathcal{L}(Q,X,Y,Z,\phi)$ theories of gravity. For this purpose, we make use of the previously developed post-Newtonian expansion of symmetric teleparallel gravity theories~\cite{Flathmann:2021khc,Hohmann:2021zel}, which we enhance by including a post-Newtonian expansion for the scalar field and a Taylor expansion for the free function defining the action, in full analogy to the case of scalar-torsion gravity~\cite{Flathmann:2019khc}. Our conventions and notation follow the textbook~\cite{Will:1993ns}.

The outline of the article is as follows. In section~\ref{sec:action}, we briefly review the field variables of scalar-nonmetricity gravity and introduce the class of theories we study in the remaining section. We briefly discuss the post-Newtonian expansion of the field equations in section~\ref{sec:ppn}. In section~\ref{sec:solution}, we solve these field equations up to the required perturbation order. The resulting PPN parameters are shown and interpreted in section~\ref{sec:param}. We end with a conclusion in section~\ref{sec:conclusion}.

\section{Field variables and their dynamics}\label{sec:action}
Before defining the action and the resulting field equations of the class of $\mathcal{L}(Q,X,Y,Z,\phi)$ scalar-nonmetricity theories, we intend to review the underlying dynamical fields. As usual in theories where nonmetricity is the mediator of gravity, the dynamical fields are a Lorentzian metric $g_{\mu\nu}$ and an affine connection $\Gamma^{\rho}{}_{\mu\nu}$. In addition we couple a dynamical scalar field $\phi$. We specify the properties of the connection by demanding vanishing torsion
\begin{equation}\label{eqn:torsion}
    T^{\rho}{}_{\mu\nu}=-2\Gamma^{\rho}{}_{[\mu\nu]}=0\,
\end{equation}
and curvature
\begin{equation}\label{eqn:curvature}
    R^{\rho}{}_{\sigma\mu\nu}=2\partial_{[\mu}\Gamma^{\rho}{}_{|\sigma|\nu]}+2\Gamma^{\rho}{}_{\lambda [\mu}\Gamma^{\lambda}{}_{|\sigma|\nu]}=0\,,
\end{equation}
whereas the covariant derivative of the metric with respect to the dynamical connection (i.e., nonmetricity) is nonzero
\begin{equation}
    Q_{\rho\mu\nu}=\nabla_{\rho}g_{\mu\nu}\neq 0\,.
\end{equation}
The combination of Eqns. \eqref{eqn:torsion} and  \eqref{eqn:curvature} leads to the form of the connection
\begin{equation}\label{eqn:connection}
    \Gamma^{\rho}{}_{\mu\nu}=\left(\Lambda^{-1}\right)^{\rho}{}_{\lambda}\partial_{\nu}\Lambda^{\lambda}{}_{\mu}\,,
\end{equation}
with $\partial_{[\mu}\Lambda^{\lambda}{}_{\nu]} = 0$. We consider an action of the form
    \begin{equation}
    S[g_{\mu\nu},\Gamma^{\rho}{}_{\mu\nu},\phi,\chi]=S_g[g_{\mu\nu},\Gamma^{\rho}{}_{\mu\nu},\phi]+ S_m[g_{\mu\nu},\chi]\,,
\end{equation}
where $\chi$ is an arbitrary set of matter field fields. The gravitational part of the action
\begin{equation}
    S_g[g_{\mu\nu},\Gamma^{\rho}{}_{\mu\nu},\phi]=\frac{1}{2\kappa^2}\int_{M}\dd^4x\sqrt{-g}\mathcal{L}(Q,X,Y,Z,\phi)\,,
\end{equation}
is a free function of the scalar field $\phi$, the nonmetricity scalar
\begin{equation}
    Q=-\frac{1}{4}Q_{\mu\nu\rho}Q^{\mu\nu\rho}+\frac{1}{2}Q_{\mu\nu\rho}Q^{\rho\nu\mu}+\frac{1}{4}Q_{\mu}Q^{\mu}-\frac{1}{2}Q_{\mu}\tilde{Q}^{\mu}\,,
\end{equation}
the kinetic term of the scalar field
\begin{equation}
    X=-\frac{1}{2}g^{\mu\nu}\partial_{\mu}\phi\partial_{\nu}\phi\,,
\end{equation}
and the derivative couplings
\begin{equation}
    Y=Q^{\mu}\partial_{\mu}\phi\,
\end{equation}
and
\begin{equation}
    Z=\tilde{Q}^{\mu}\partial_{\mu}\phi\,,
\end{equation}
which couple the scalar field to the two independent contractions of the nonmetricity tensor
\begin{equation}
Q^{\mu}=Q^{\mu\rho}{}_{\rho}\,,\quad \tilde{Q}^{\mu}=Q_{\rho}{}^{\rho\mu}\,.
\end{equation}
By varying the matter action $S_m$ with respect to the metric
\begin{equation}
\delta S_m=-\frac{1}{2}\int_M\Theta^{\mu\nu}\delta g_{\mu\nu}\sqrt{-g}\dd^4x
\end{equation}
we obtain the energy momentum tensor $\Theta^{\mu\nu}$. The full variation of the action $S$ with respect to the metric then leads to the field equations
\begin{align}
   0=E_{\mu\nu}=&-\mathcal{L}g_{\mu\nu}+\lc{\nabla}_{\rho}\left(\mathcal{L}_QP^{\rho}{}_{\mu\nu}\right)+\frac{1}{2}g_{\mu\nu}g^{\rho\sigma}\lc{\nabla}_{\rho}\left(\mathcal{L}_Y\partial_{\rho}\phi\right)+\frac{1}{2}\lc{\nabla}_{(\mu}\left(\mathcal{L}_Z\partial_{\nu)}\phi\right)\nonumber\\
    &-\frac{1}{2}\mathcal{L}_Q\left(2Q^{\rho}{}_{\mu\sigma}\left[Q_{\rho\nu}{}^{\sigma}-Q^{\sigma}{}_{\rho\nu}\right]-Q_{\mu}{}^{\rho\sigma}Q_{\nu\rho\sigma}+Q_{\rho}\left[2Q_{(\mu\nu)}{}^{\rho}-Q^{\rho}{}_{\mu\nu}\right]\right)\nonumber\\
    &-\mathcal{L}_X\partial_{\mu}\phi\partial_{\nu}\phi+2\mathcal{L}_YQ_{(\mu}\partial_{\nu)}\phi-\mathcal{L}_Z\left(Q^{\rho}{}_{\mu\nu}\partial_{\rho}\phi-2Q_{(\mu\nu)}{}^{\rho}\partial_{\rho}\phi-Q_{(\mu}\partial_{\nu)}\phi\right)\nonumber\\
    &-\kappa^2\Theta_{\mu\nu}\,
\end{align}
and similarly a variation with respect to the scalar field leads to the scalar field equation
\begin{equation}
    0 = E_{\phi}=\lc{\nabla}^{\mu}\left(\mathcal{L}_YQ_{\mu}+\mathcal{L}_Z\tilde{Q}_{\mu}-\mathcal{L}_Y\partial_{\mu}\phi\right)-\mathcal{L}_{\phi}\,.
\end{equation}
Note that $\mL_{Q,X,Y,Z,\phi}$ is the derivative of the free function $\mL$ with respect to $Q \,,X\,, Y\,, Z$ and $\phi$, respectively. We finally remark that another field equation can be obtained by variation of the action with respect to the flat, symmetric affine connection; however this field equation is fully determined from the previously displayed equations through the Bianchi identities, and thus redundant~\cite{Hohmann:2021fpr}. We therefore omit it here for brevity and show only the independent equations, which we will solve in the following sections.

\section{Post-Newtonian approximation}\label{sec:ppn}
In order to compare this family of theories with observations, we employ the parametrized post-Newtonian (PPN) formalism in its form detailed in the textbook \cite{Will:1993ns}.\footnote{A slightly different formulation, with different conventions regarding the definition of the potentials, is used in the new edition \cite{Will:2018bme}.} First, we give some general remarks on the PPN formalism and then, we review how to perform the post-Newtonian expansion of the dynamical connection. We start with the description of the matter part of the field equation. As usual we assume a perfect fluid of the form
\begin{equation}\label{eqn:tmunu}
\Theta^{\mu\nu} = (\rho + \rho\Pi + p)u^{\mu}u^{\nu} + pg^{\mu\nu}\,,
\end{equation}
with $\rho$, $\Pi$, $p$ and $u^{\mu}$ being the rest energy density, specific internal energy, pressure and four velocity, respectively. We further assume a normalization of $u^{\mu}u^{\nu}g_{\mu\nu}=-1$ for the four velocity and compared to the speed of light $c=1$ the velocity of the matter $v^i=u^i/u^0$ in a given reference frame is assumed to be small. Next, we perform a perturbative expansion in orders of the velocity $\mathcal{O}(n) \propto |\vec{v}|^n$. This has to be done for the metric $g_{\mu\nu}$, the matter fields $\Theta^{\mu\nu}$, the dynamical connection $\Gamma^{\rho}{}_{\mu\nu}$, the scalar field $\phi$ and the free function $\mathcal{L}$. The metric $g_{\mu\nu}$ will be expanded around the flat Minkowski metric $\eta_{\mu\nu}=\mathrm{diag}(-1,1,1,1)$
\begin{equation}\label{eqn:metricperturb}
g_{\mu\nu}=\eta_{\mu\nu}+h_{\mu\nu}=\eta_{\mu\nu}+\order{h}{2}_{\mu\nu}+\order{h}{3}_{\mu\nu}+\order{h}{4}_{\mu\nu}+\mathcal{O}(5)\,.
\end{equation}
As a consequence, the energy-momentum tensor reads as
\begin{subequations}\label{eqn:energymomentum}
\begin{align}
\Theta_{00} &= \rho\left(1 + \Pi + v^2 - \order{h}{2}_{00}\right) + \mathcal{O}(6)\,,\\
\Theta_{0j} &= -\rho v_j + \mathcal{O}(5)\,,\\
\Theta_{ij} &= \rho v_iv_j + p\delta_{ij} + \mathcal{O}(6)\,.
\end{align}
\end{subequations}
Here, we used the standard assumptions for the orders of the matter fields. Next, we make use of the form of the connection in Eqn.~\eqref{eqn:connection}. As developed in \cite{Flathmann:2021khc}, we expand the coordinates around the coordinates of the coincident gauge up to quadratic orders of the generators of a ``knight diffeomorphism''
\begin{equation}
x'^{\mu} = x^{\mu} + \xi^{\mu} + \frac{1}{2}\xi^{\nu}\partial_{\nu}\xi^{\mu}\,,
\end{equation}
with which the connection can be written as
\begin{equation}
\Gamma^{\rho}{}_{\mu\nu} = \partial_{\mu}\partial_{\nu}\xi^{\rho} + \frac{1}{2}\left(\xi^{\sigma}\partial_{\mu}\partial_{\nu}\partial_{\sigma}\xi^{\rho} + 2\partial_{(\mu}\xi^{\sigma}\partial_{\nu)}\partial_{\sigma}\xi^{\rho} - \partial_{\mu}\partial_{\nu}\xi^{\sigma}\partial_{\sigma}\xi^{\rho}\right)\,.
\end{equation}
Now, we expand $\xi^{\mu}$ similar to the metric in post-Newtonian orders
\begin{equation}
\xi^{\alpha}=\order{\xi}{2}^{\alpha}+\order{\xi}{3}^{\alpha}+\order{\xi}{4}^{\alpha}+\mathcal{O}\left(5\right)\,.
\end{equation}
Furthermore, we expand the scalar field $\phi$ around its cosmological background value $\Phi$, which we assume to be constant
\begin{equation}
    \phi=\Phi+\psi=\Phi+\order{\psi}{1}+\order{\psi}{2}+\order{\psi}{3}+\order{\psi}{4}\,.
\end{equation}
The components of the dynamical fields, we have to calculate are
\begin{equation}
\order{h}{2}_{00}\,,\order{h}{2}_{ij}\,,\order{h}{3}_{i0},\order{h}{4}_{00}\,,\order{\xi}{2}^{i}\,,\order{\xi}{3}^{0}\,,\order{\xi}{4}^{i}\,, \order{\psi}{2}\,.
\end{equation}
Lastly, we have to deal with the free function $\mathcal{L}$ in the action. For this we perform a Taylor expansion, where we assume the Taylor coefficients to be of velocity order $\mathcal{O}(0)$:
\begin{align}\label{eqn:LXYphiTaylor}
\mathcal{L} &= l_0+l_\phi\psi+\frac{1}{2}l_{\phi\phi}\psi^2+l_Q Q+l_XX+l_YY+l_ZZ \,, \nonumber \\
\mathcal{L}_Q &= l_Q+l_{T\phi}\psi+\frac{1}{2}l_{Q\phi\phi}\psi^2+l_{QX}X+l_{QY}Y+l_{QZ}Z+l_{QQ}Q \,, \nonumber \\
\mathcal{L}_X &= l_X+l_{X\phi}\psi+\frac{1}{2}l_{X\phi\phi}\psi^2+l_{QX}Q+l_{XY}Y+l_{XZ}Z+l_{XX}X \,, \nonumber \\
\mathcal{L}_Y &= l_Y+l_{Y\phi}\psi+\frac{1}{2}l_{Y\phi\phi}\psi^2+l_{QY}Q+l_{XY}X+l_{YZ}Z+l_{YY}Y \,, \nonumber \\
\mathcal{L}_Z &= l_Y+l_{Z\phi}\psi+\frac{1}{2}l_{Z\phi\phi}\psi^2+l_{QZ}Q+l_{XZ}X+l_{YZ}Z+l_{ZZ}Z \,, \nonumber \\
\mathcal{L}_\phi &= l_\phi +l_{\phi\phi}\psi+l_{Q\phi}T+l_{X\phi}X+l_{Y\phi}Y+l_{Z\phi}Z+\frac{1}{2}l_{\phi\phi\phi}\psi^2 \,.
\end{align}
By combining all perturbative expansions of this section, we can now calculate and solve the field equations in the next section.

\section{Solving the field equations}\label{sec:solution}
We now apply the post-Newtonian expansion displayed in the previous section to the class of symmetric teleparallel gravity theories outlined in section~\ref{sec:action}, in order to derive and solve the post-Newtonian field equations. We proceed in ascending velocity orders; the zeroth, second, third and fourth velocity order is discussed in sections~\ref{ssec:order0}, \ref{ssec:order2}, \ref{ssec:order3} and~\ref{ssec:order4}, respectively. Calculations have been performed using \emph{xPPN}~\cite{Hohmann:2020muq}. Alternatively, one could make use of the gauge-invariant approach to the PPN formalism~\cite{Hohmann:2019qgo,Hohmann:2021zel}; the resulting equations for the constant coefficients we obtain are identical.

\subsection{Zeroth order and assumption}\label{ssec:order0}
First of all for simplicity we assume a massless scalar field in order to avoid solutions in terms of Yukawa type potentials. This can be achieved by assuming both $l_{\phi\phi}$ and $l_{\phi\phi\phi}=0$. The zeroth order equations (calculated by inserting $g_{\mu\nu}=\eta_{\mu\nu}$ and $\phi=\Phi$) read
\begin{equation}
0 = l_0\eta_{\mu\nu}\,, \quad
0 = l_{\phi}\,.
\end{equation}
Therefore, the perturbed metric is given in the standard PPN form if and only if $l_0=l_{\phi}=0$. For the remainder of this article, we will use these assumptions.

\subsection{Second order}\label{ssec:order2}
With the assumptions and the solutions of the zeroth order field equations, we can now calculate the second order field and scalar field equations. The only nonvanishing components read
\begin{align}
\order{E}{2}_{00} &=\kappa^2\rho-\frac{1}{2}l_Q\left(\partial_j\partial_i\order{h}{2}^{ij}-\triangle\order{h}{2}^i{}_i\right)+l_Y\triangle\order{\phi}{2}=0\,,\nonumber\\
\order{E}{2}_{ij} &= \frac{1}{2}l_Q\left(-\partial_j\partial_i\order{h}{2}_{00}+\partial_j\partial_i\order{h}{2}^k{}_{k}+2\partial_{k}\partial_{(i}\order{h}{2}^k{}_{j)}+\triangle\order{h}{2}_{ij}+\delta_{ij}\left[\triangle\order{h}{2}_{00}+\partial_k\partial_l\order{h}{2}^{kl}-\triangle\order{h}{2}^k{}_{k}\right]\right)-\delta_{ij} l_Y\triangle\order{\phi}{2}-l_Z\partial_j\partial_i\order{\phi}{2}=0\,,\nonumber\\
\order{E}{2}_{\phi} &= l_X\triangle\order{\phi}{2}+l_Y\triangle\left(\order{h}{2}_{00}-\order{h}{2}^{k}{}_k+2\partial_k\order{\xi}{2}^k\right)+l_Z\left(-\partial_k\partial_i\order{h}{2}^{ij}+2\triangle\partial_k\order{\xi}{2}^k\right)=0\,.
\end{align}
These three equations can be solved with the ansatz
\begin{equation}
\order{h}{2}_{00} =a_1U\,, \quad
\order{h}{2}_{ij} = a_2\delta_{ij}U\,, \quad
\order{\xi}{2}^{i} = a_3\partial^i\chi\,, \quad
\order{\phi}{2} =a_4U\,.
\end{equation}
Here, $U$ and $\chi$ are the usual PPN potentials, which are defined by the relations $\triangle\chi=-2U$ and \(\triangle U = -4\pi\rho\). Inserting this ansatz into the field equations leads to a system of algebraic equations for the coefficients $a_i$. In order to determine the most general solution to this system, one must distinguish different cases. The solution for the generic case is given by
\begin{equation}
a_1 =\frac{\kappa^2}{4\pi l_Q} \,, \quad
a_2 =\frac{\kappa^2}{4\pi l_Q} \,, \quad
a_3 = -\frac{\kappa^2\left(l_Z+2l_Y\right)}{16\pi\left(l_Z+l_Y\right)l_Q}\,, \quad
a_4 = 0 \,,
\end{equation}
and is valid if and only if the denominator \((l_Y + l_Z)l_Q\) is non-vanishing. Otherwise, the system is degenerate and one must further distinguish between two cases. For \(l_Q = 0\), one cannot solve for \(a_1\). Since this component is required for the Newtonian limit of the theory, as it governs the contribution of the Newtonian potential, we conclude that this case is not physically viable, and henceforth assume \(l_Q \neq 0\). Further assuming \(l_Y + l_Z = 0\), one cannot solve for \(a_3\), as it cancels from the algebraic equations. For the remaining coefficients one obtains the solution
\begin{equation}
a_1 = \frac{\kappa^2}{4\pi l_Q}\frac{4l_Y^2 + l_Xl_Q}{3l_Y^2 + l_Xl_Q}\,, \quad
a_2 = \frac{\kappa^2}{4\pi l_Q}\frac{2l_Y^2 + l_Xl_Q}{3l_Y^2 + l_Xl_Q}\,, \quad
a_4 = \frac{\kappa^2}{4\pi}\frac{l_Y}{3l_Y^2 + l_Xl_Q}\,,
\end{equation}
provided that \(3l_Y^2 + l_Xl_Q \neq 0\). Otherwise, if \(3l_Y^2 + l_Xl_Q = 0\) with \(l_Y \neq 0\), one finds that the system does not possess any solution with non-vanishing matter content.  Finally, if \(l_X = l_Y = 0\), one obtains the solution
\begin{equation}
a_1 = a_2 = \frac{\kappa^2}{4\pi l_Q}\,.
\end{equation}
This contains general relativity as a special case.

\subsection{Third order}\label{ssec:order3}
At the third velocity order, the only non-trivial field equation is given by
\begin{equation}
\order{E}{3}_{0i} = -\kappa^2\rho v_i - l_Z\partial_0\partial_i\order{\phi}{2} + l_Q\left(\partial_0\partial_{[i}\order{h}{2}_{j]}{}^j + \partial^j\partial_{[j}\order{h}{2}_{i]0}\right)\,.
\end{equation}
Note in particular that the third order connection component \(\order{\xi}{3}^0\) does not enter these equations and thus remains undetermined, so that we can solve for the metric perturbation \(\order{h}{3}_{0i}\). This can be done by using the ansatz
\begin{equation}
\order{h}{3}_{0i} = a_5V_i + a_6W_i\,,
\end{equation}
where \(V_i\) and \(W_i\) denote the standard PPN potentials defined in~\cite{Will:1993ns}. Further, we need to substitute the second order perturbations found in the previous section for the different non-degenerate and degenerate cases. It is remarkable that in all cases this procedure leads to the same coefficient equation
\begin{equation}
\kappa^2 + 2\pi l_Q(a_5 + a_6) = 0\,,
\end{equation}
and hence we have the solution
\begin{equation}
a_5 + a_6 = -\frac{\kappa^2}{2\pi l_Q}\,,
\end{equation}
while their difference is not determined by the third order field equations. The latter is an expected result, as it reflects the invariance of the theory and its post-Newtonian limit under (infinitesimal) diffeomorphisms.

\subsection{Fourth order}\label{ssec:order4}
We finally come to the fourth velocity order. We will not display the full perturbative expansion of the field equations here, as it turns out to be lengthy, and restrict ourselves to presenting the steps which are necessary to obtain the solution, starting with the non-degenerate case \(l_Q(l_Y + l_Z) \neq 0\). In this case we find that the fourth order field equations contain besides the component \(\order{h}{4}_{00}\), which we need to solve for in order to determine the PPN parameters, also the fourth order components \(\order{h}{4}_{ij}\), \(\order{\xi}{4}^i\) and \(\order{\phi}{4}\). It turns out that these can be eliminated from the fourth order equations by considering the linear combination
\begin{equation}
(2l_Y + l_Z)\partial_i\partial_j\order{E}{4}^{ij} - (l_Y + l_Z)\triangle(\order{E}{4}_{00} + \order{E}{4}_i{}^i) = 0\,.
\end{equation}
To solve this equation, we make an ansatz of the form
\begin{equation}\label{eqn:metans4}
\order{h}{4}_{00} = a_7U^2 + a_8\Phi_1 + a_9\Phi_2 + a_{10}\Phi_3 + a_{11}\Phi_4 + a_{12}\Phi_W + a_{13}\mathcal{A}\,,
\end{equation}
once again referring to~\cite{Will:1993ns} for the definition of the appearing PPN potentials. In addition to the coefficients \(a_7, \ldots, a_{13}\) in this ansatz, we also need to determine the linear combination \(a_5 - a_6\) from the third order ansatz, which is left undetermined in the third order equations. By extracting the coefficients of the independent matter terms in the fourth order equations, we find that they indeed possess a unique solution for these coefficients, which reads
\begin{equation}\label{eqn:coeff4nondeg}
a_5 - a_6 = -\frac{3\kappa^2}{8\pi l_Q}\,, \;
a_7 = -\frac{\kappa^4}{32\pi^2l_Q^2}\,, \;
a_8 = \frac{\kappa^2}{2\pi l_Q}\,, \;
a_9 = \frac{\kappa^4}{16\pi^2l_Q^2}\,, \;
a_{10} =  \frac{\kappa^2}{4\pi l_Q}\,, \;
a_{11} =  \frac{3\kappa^2}{4\pi l_Q}\,, \;
a_{12} = a_{13} = 0\,.
\end{equation}
In the degenerate case \(l_Y + l_Z = 0\) and \(3l_Y^2 + l_Xl_Q \neq 0\), the fourth order connection component \(\order{\xi}{4}^i\) does not enter the field equations. One can isolate the component \(\order{h}{4}_{00}\) from the remaining fourth order components by taking the linear combination
\begin{equation}
(4l_Y^2 + l_Xl_Q)\order{E}{4}_{00} + (2l_Y^2 + l_Xl_Q)\order{E}{4}_i{}^i + l_Yl_Q\order{E}{4}_{\phi} = 0\,.
\end{equation}
Using again the ansatz~\eqref{eqn:metans4}, we now find the solution
\begin{gather}
a_5 - a_6 = -\frac{\kappa^2}{8\pi l_Q}\frac{8l_Y^2 + 3l_Xl_Q}{3l_Y^2 + l_Xl_Q}\,, \quad
a_{10} = \frac{\kappa^2}{4\pi l_Q}\frac{4l_Y^2 + l_Xl_Q}{3l_Y^2 + l_Xl_Q}\,, \quad
a_{11} = \frac{3\kappa^2}{4\pi l_Q}\frac{2l_Y^2 + l_Xl_Q}{3l_Y^2 + l_Xl_Q}\,, \quad
a_8 = \frac{\kappa^2}{2\pi l_Q}\,,\nonumber\\
a_{12} = a_{13} = 0\,, \quad
a_9 - 2a_7 = \frac{\kappa^4}{16\pi^2l_Q^2}\frac{20l_Y^4 + 13l_Xl_Y^2l_Q + 2l_X^2l_Q^2}{(3l_Y^2 + l_Xl_Q)^2}\,, \quad
a_9 + 2a_7 = -\frac{\kappa^4l_Y}{16\pi^2l_Q^2}\label{eqn:coeff4deg}\\
\times\frac{24l_Y^5 + 2l_Y^3l_Q[7l_X - 2(l_{Y\phi} + l_{Z\phi})] + l_Yl_Q^2[2l_X(l_X - 2l_{Y\phi} - l_{Z\phi}) + l_Yl_{X\phi}] + (2l_Y^2 + l_Xl_Q)(6l_Y^2 + l_Xl_Q)l_{Q\phi}}{(3l_Y^2 + l_Xl_Q)^3}\,.\nonumber
\end{gather}
Finally, for \(l_X = l_Y = l_Z = 0\) we find again the same solution~\eqref{eqn:coeff4nondeg} as for the non-degenerate case. Hence, we have determined all possible solutions for the fourth order, without introducing any further distinction between different cases beyond the one introduced at the second order.

\section{PPN parameters}\label{sec:param}
\begin{figure}[htbp]
\begin{tikzpicture}[decision/.style={draw,rectangle,fill=black!10!white,outer sep=2pt},result/.style={draw,rectangle,rounded corners,outer sep=2pt},every edge/.style={draw,-Stealth},good/.style={fill=green!25!white},maybe/.style={fill=yellow!25!white},noppn/.style={fill=red!25!white},degen/.style={fill=magenta!25!white}]
\node[decision] (l0lp) at (0,0) {$l_0, l_{\phi}$};
\node[decision] (lpplppp) at (0,-2) {$l_{\phi\phi}, l_{\phi\phi\phi}$};
\node[decision] (lq) at (0,-4) {$l_Q$};
\node[decision] (lylz) at (0,-6) {$l_Y + l_Z$};
\node[decision] (ly) at (0,-8) {$l_Y$};
\node[decision] (lxlylq) at (0,-10) {$3l_Y^2 + l_Xl_Q$};
\node[decision] (lxlylq2) at (0,-12) {$4l_Y^2 + l_Xl_Q$};
\node[result,noppn] (nominkbg) at (4,0) {no Minkowski background};
\node[result,maybe] (nonconst) at (4,-2) {non-constant PPN parameters};
\node[result,degen] (nonewton) at (4,-8) {no Newtonian limit};
\node[result,good] (ppngood) at (-4,-8) {$\beta = \gamma = 1$};
\node[result,maybe] (ppnbad) at (-4,-12) {$\beta \neq 1, \gamma \neq 1$};
\node[result,degen] (nomatter) at (-4,-10) {no matter allowed};
\path (0,1) edge[ultra thick] (l0lp);
\path (l0lp) edge node[above, near start] {$\neq 0$} (nominkbg);
\path (l0lp) edge[ultra thick] node[left, near start] {$= 0$} (lpplppp);
\path (lpplppp) edge node[above, near start] {$\neq 0$} (nonconst);
\path (lpplppp) edge[ultra thick] node[left, near start] {$= 0$} (lq);
\path (lq) edge node[above right, near start] {$= 0$} (nonewton);
\path (lq) edge[ultra thick] node[left, near start] {$\neq 0$} (lylz);
\path (lylz) edge node[above left, near start] {$\neq 0$} (ppngood);
\path (lylz) edge[ultra thick] node[left, near start] {$= 0$} (ly);
\path (ly) edge node[left, near start] {$\neq 0$} (lxlylq);
\path (ly) edge[ultra thick] node[below, near start] {$= 0$} (ppngood);
\path (lxlylq) edge node[left, near start] {$\neq 0$} (lxlylq2);
\path (lxlylq) edge node[above, near start] {$= 0$} (nomatter);
\path (lxlylq2) edge node[above, near start] {$\neq 0$} (ppnbad);
\path (lxlylq2) edge node[below right, near start] {$= 0$} (nonewton);
\end{tikzpicture}
\caption{Full classification of $\mathcal{L}(Q, X, Y, Z, \phi)$ theories. The path highlighted by thick arrows corresponds to STEGR. Theories with \(\beta = \gamma = 1\) are in full agreement with observations. Theories with deviating, but constant PPN parameters receive bounds on their parameters, and are still in agreement if these bounds are met. Theories with massive scalar fields possess distance-dependent PPN parameters and need a more thorough treatment. Other classes of theories are either pathological or need an extension to the standard PPN formalism.}
\label{fig:class}
\end{figure}
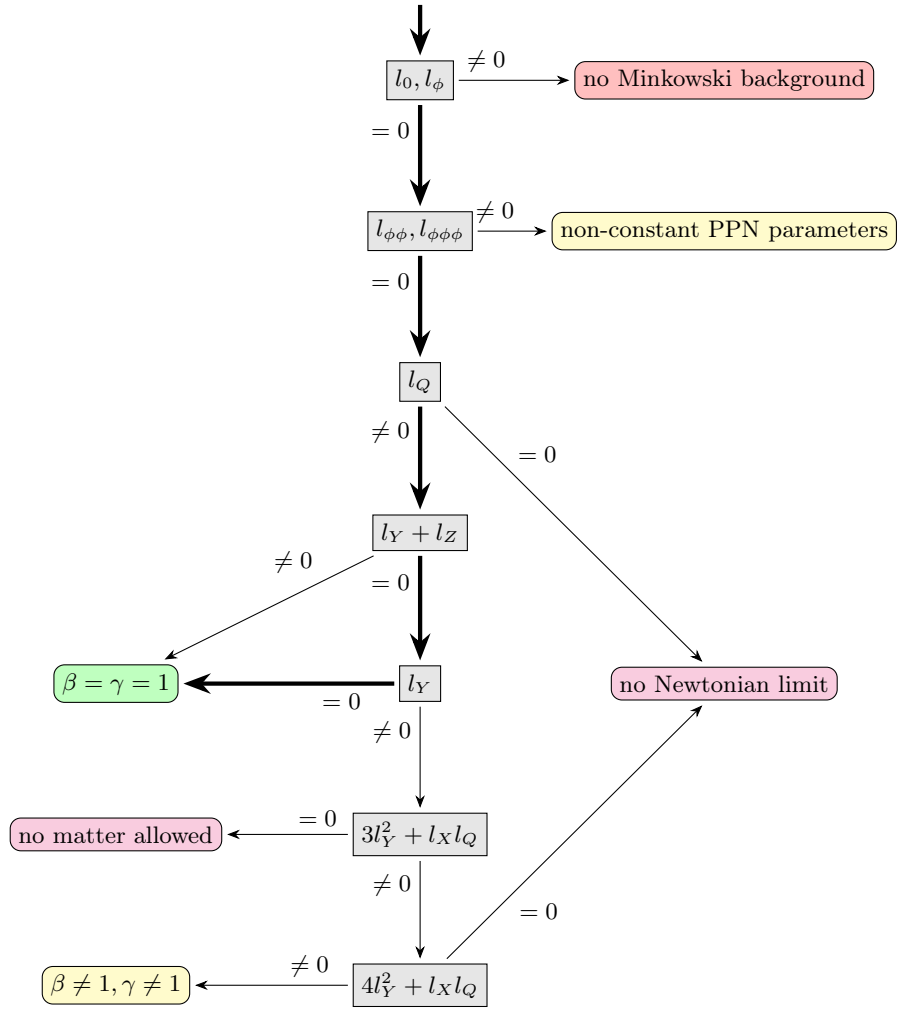

By comparing the solution for the metric perturbation components we have derived in the preceding section to their standard PPN form~\cite{Will:1993ns,Will:2014kxa}, we are now able to obtain the values of the PPN parameters for the different subclasses of scalar-nonmetricity theories we considered. We start by recalling that the theories we study here are restricted by the conditions \(l_0 = l_{\phi} = 0\) in order to possess a Minkowski background solution, \(l_{\phi\phi} = l_{\phi\phi\phi} = 0\) for a massless scalar field and \(l_Q \neq 0\) to obtain a well-defined Newtonian limit. We find that after imposing these conditions, the most generic class of theories satisfying \(l_Y + l_Z \neq 0\) exhibits the PPN parameters
\begin{equation}\label{eqn:grppnval}
\beta = \gamma = 1\,, \quad
\alpha_1 = \alpha_2 = \alpha_3 = \zeta_1 = \zeta_2 = \zeta_3 = \zeta_4 = \xi = 0\,,
\end{equation}
and thus fully agrees with the PPN parameters of general relativity. Potential deviations from these values are encountered only in the subclass \(l_Y + l_Z = 0\). Within this subclass, we found that theories which in addition satisfy \(l_X = l_Y = 0\), so that the scalar field is minimally coupled to nonmetricity at the linear order, again yield the same PPN parameters~\eqref{eqn:grppnval}. For theories with \(3l_Y^2 + l_Xl_Q \neq 0\) we obtain the PPN parameters
\begin{subequations}
\begin{align}
\gamma &= 1 - \frac{2l_Y^2}{4l_Y^2 + l_Xl_Q}\,,\\
\beta &= 1 - \frac{l_Y\{12l_Y^5 + l_Y^3l_Q[7l_X + 4(l_{Y\phi} + l_{Z\phi})] + l_Yl_Q^2[l_X(l_X + 4l_{Y\phi} + 2l_{Z\phi}) - l_Yl_{X\phi}] - (2l_Y^2 + l_Xl_Q)(6l_Y^2 + l_Xl_Q)l_{Q\phi}\}}{2(3l_Y^2 + l_Xl_Q)(4l_Y^2 + l_Xl_Q)^2}\,,
\end{align}
\end{subequations}
while the remaining parameters vanish again, indicating that the theory is fully conservative, i.e., there are no preferred-frame or preferred-location effects or violation of total energy-momentum conservation. Taking a closer look at this result, one finds that even in this class there is another subclass given by \(l_Y = 0\) corresponding to a minimally coupled scalar field at the linear perturbation order which leads to the general relativity values~\eqref{eqn:grppnval}. Further, we find that for \(4l_Y^2 + l_Xl_Q = 0\) the PPN parameters \(\beta\) and \(\gamma\) diverge. This is due to the fact that in this case the contribution of the Newtonian potential \(U\) to the perturbation \(\order{h}{2}_{00}\) vanishes. Hence, also in these theories no physically meaningful Newtonian limit is obtained. We summarize our findings by listing all cases we studied and their corresponding results in diagram~\ref{fig:class}.

We conclude from our findings that the generic class of scalar-nonmetricity theories of gravity with a massless scalar field possesses the same PPN parameters as general relativity, and is therefore indistinguishable from the latter, and passes all solar system tests. Deviations for the parameters \(\beta\) and \(\gamma\) are found only for a particular subclass, which contains the scalar-nonmetricity equivalent of scalar-curvature gravity as a special case~\cite{Jarv:2018bgs}. In this case, solar system observations give bounds on the Taylor coefficients of the Lagrangian function, and so also this subclass contains theories passing the solar system tests. Another possibility, which remains to be studied, is theories in which the scalar field is massive, and in which its contribution to the post-Newtonian limit depends on the distance to the gravitating body. However, this study exceeds the scope of this article.

\section{Conclusion}\label{sec:conclusion}
We have studied the post-Newtonian limit of a general class of scalar-nonmetricity theories of gravity and calculated their PPN parameters for the case of a massless scalar field. Our results show that generically a scalar field which is non-minimally coupled to nonmetricity is suppressed in the post-Newtonian limit and does not contribute to the post-Newtonian dynamics, so that the PPN parameters agree with those of general relativity. Deviations for the PPN parameters \(\beta\) and \(\gamma\) are found only for a specific subclass of theories which are distinguished by their coupling between the scalar field and nonmetricity. Further, we find that also in this case all other PPN parameters agree with those of general relativity. Hence, we find that the theories are fully conservative and do not possess any violation of local position invariance, local Lorentz invariance or total energy-momentum conservation.

The most remarkable result of our work is the suppression of the scalar field in the post-Newtonian limit if it is non-minimally coupled to the nonmetricity via any other linear combination than the mixed trace \(Q^{\nu}{}_{\nu\mu} - Q_{\mu\nu}{}^{\nu}\) at the linear order. This particular coupling term also appears in other results. Coupling to this term restores the conformal invariance of scalar-nonmetricity theories when this is broken through a non-minimal coupling to the nonmetricity scalar \(Q\)~\cite{Jarv:2018bgs}. Also in cosmology the behavior of theories coupled to this term only differs qualitatively from the generic coupling~\cite{Hohmann:2021ast}. The fact that their PPN parameters are exactly identical to those of general relativity, while allowing for a richer cosmological dynamics, motivates further studies of this class of theories and their implications for observations in cosmology and gravitational waves, where higher order effects beyond the PPN parameters become relevant due to the strong gravity present at the gravitational wave source.

Another possible line of future investigation is to allow for a massive scalar field and study the resulting PPN parameters, in analogy to previous works on scalar-curvature~\cite{Hohmann:2013rba,Hohmann:2015kra,Hohmann:2017qje} and scalar-torsion theories of gravity~\cite{Emtsova:2019qsl}. A natural question is whether the aforementioned suppression of the scalar field is present also in this case, which would significantly simplify the post-Newtonian limit compared to the case of a non-vanishing scalar field contribution. In the latter case, the PPN parameters are no longer constant, but attain a dependence on the distance to the gravitating source, which is in general highly non-trivial.

\begin{acknowledgments}
KF gratefully acknowledges support by the DFG within the Research Training Group \textit{Models of Gravity}. MH gratefully acknowledges the full financial support by the Estonian Research Council through the Personal Research Funding project PRG356 and by the European Regional Development Fund through the Center of Excellence TK133 ``The Dark Side of the Universe''. This article is based upon work from COST Action QGMM (CA18108), supported by COST (European Cooperation in Science and Technology).
\end{acknowledgments}

\bibliography{teleppn}

\end{document}